\documentclass[11pt]{article}
\usepackage{amssymb}
\usepackage{amsmath}
\usepackage{rsfs}    

\newcommand{\bra}{\langle}
\newcommand{\ket}{\rangle}

\begin{document} 
\baselineskip 6.0mm 
\begin{flushright}
submitted on November 4, 2014
\\
modified on April 6, 2015
\end{flushright}
\vspace*{1mm}
\begin{center}
\baselineskip 8.0mm 
{\bf \LARGE
Uncertainty relation between angle and orbital angular momentum: 
interference effect in electron vortex beams
}\footnote{
This manuscript is written as a part of the proceedings of the workshop,
Wave dynamics in low-dimensional branched structures,
held during September 23-24, 2014 in Tashkent, Uzbekistan.
It is to be published in the journal, Nanosystems: Physics, Chemistry, Mathematics.}
\vspace{4mm} \\
\baselineskip 6.0mm 
Shogo Tanimura
\\
{\it
Department of Complex Systems Science,
Graduate School of Information Science,
Nagoya University,
Nagoya 464-8601, Japan
}
\\
{\tt e-mail: tanimura[AT]is.nagoya-u.ac.jp}
\vspace{10mm}
\\
\begin{minipage}{110mm}
Abstract: 
The uncertainty relation
between angle and orbital angular momentum
had not been formulated in a similar form as
the uncertainty relation between position and linear momentum
because the angle variable is not represented
by a quantum mechanical self-adjoint operator.
Instead of the angle variable operator,
we introduce the complex position operator 
$ \hat{Z} = \hat{x}+i \hat{y} $
and interpret the order parameter
$ \mu = 
\bra \hat{Z} \ket / \sqrt{ \bra \hat{Z}^\dagger \hat{Z} \ket} $
as a measure of certainty of the angle distribution.
We prove the relation 
between the uncertainty of angular momentum
and the angle order parameter.
We prove also its generalizations
and discuss experimental methods for testing these relations.
\end{minipage}
\end{center}
\vspace{6mm}
\vspace{4mm}
\baselineskip 5.5mm 

\section{Introduction}
The uncertainty relations elucidate
the difference between classical physics and quantum physics.
In classical physics, accuracy of measurement is not limited in principle
and it is assumed that 
any observables can be measured simultaneously and precisely.
However, in quantum physics, 
the accuracy of simultaneous measurements of two observables
is limited by the uncertainty relation.

Originally,
Heisenberg \cite{Heisenberg1927} formulated the uncertainty relation 
between position $ Q $ and linear momentum $ P $ as
\begin{equation}
	\Delta Q \, \Delta P \, \gtrsim \, h
\end{equation}
with the Planck constant $ h $.
He deduced this relation via a Gedankenexperiment.
Later, 
Weyl, Kennard, and Robertson \cite{Robertson1929}
gave a rigorous proof of this relation.
In the context of quantum mechanics
the position is represented by a self-adjoint operator $ \hat{Q} $
and the uncertainty of the position is defined as the variance
\begin{equation}
	( \Delta Q )^2
	:=
	\Big\bra \psi \Big|
	\Big( \hat{Q} - \bra \psi | \hat{Q} | \psi \ket \Big)^2
	\Big| \psi \Big\ket
	=
	\bra \psi | \hat{Q}^2 | \psi \ket
	- \bra \psi | \hat{Q} | \psi \ket^2
	\label{Delta Q}
\end{equation}
for a normalized state vector $ | \psi \ket $.
The uncertainty $ \Delta P $ of momentum is defined in a similar way.

It is natural to expect a similar relation 
\begin{equation}
	\Delta \phi \, \Delta L \, \gtrsim \, h
\end{equation}
holds for the angle $ \phi $ and the angular momentum $ L $
as shown in the textbook \cite{Schiff1968}.
However, in a plane, 
the coordinate values $ \{ \phi + 2 \pi n \} $
with any integer $ n $ represent
the same point as $ \phi $ indicates.
In other words, the angle variable $ \phi $ is a multivalued function.
In quantum mechanics,
the spectrum of a self-adjoint operator should have 
one-to-one correspondence
with the values of an observable.
Hence, there is no self-adjoint operator $ \hat{\phi} $ 
corresponding to the multivalued angle variable $ \phi $.
Therefore, the angle uncertainty $ \Delta \phi $
cannot be defined as the position uncertainty $ \Delta Q $ was defined.

The uncertainty relation between angle and 
orbital angular momentum 
is a longstanding issue in physics.
Many people have proposed other definitions of the angle uncertainty
and have formulated several versions of the uncertainty relation 
between angle and angular momentum \cite{Judge1963}-\cite{Tanimura1993}.
However, 
most of them treat a particle moving on a one-dimensional circle.
They did not consider a particle moving in two- or three-dimensional spaces.
Thus we do not yet have 
an angle-angular momentum uncertainty relation 
that is applicable for a realistic situation.

In this paper,
we introduce the moment of position distribution in a plane,
which is an arbitrary two-dimensional subspace in the configuration space
of the particle.
We propose to use the moment of position 
as an indicator of certainty or bias of angle distribution.
Main results of this work are the inequalities
(\ref{result1}),
(\ref{result2}),
(\ref{result3}),
which represent
the uncertainty relation 
between the moments of position and the orbital angular momentum.
Our results are applicable for a particle moving 
in the configuration space 
having more than two dimensions.

\section{Robertson inequality}
The Robertson inequality \cite{Robertson1929}
is one of formulations of general uncertainty relations.
The Robertson inequality has a clear meaning
and it is applicable to any kind of observables.
Hence it is regarded as the universal formulation of uncertainty relations.
Although the Robertson inequality is well known
and its proof is rather simple,
here we write its derivation to make a comparison with
our uncertainty relation of the angle and angular momentum,
which is derived in the next section.

For any vectors $ | \alpha \ket $ and $ | \beta \ket $
of a Hilbert space $ {\mathscr H} $,
the Schwarz inequality 
\begin{equation}
	\bra \alpha | \alpha \ket
	\bra \beta  | \beta \ket
	\ge
	\Big| \bra \alpha | \beta \ket \Big|^2
	\label{Schwarz}
\end{equation}
holds.
The equality holds if and only if 
the two vectors $ | \alpha \ket $ and $ | \beta \ket $ are linearly dependent.
Let $ \psi \in {\mathscr H} $
be an arbitrary normalized vector satisfying $ \bra \psi | \psi \ket = 1 $.
For self-adjoint operators $ \hat{A} $ and $ \hat{B} $ on $ {\mathscr H} $,
we set
\begin{eqnarray}
&&	\bra \hat{A} \ket := \bra \psi | \hat{A} | \psi \ket,
	\\
&&	\Delta \hat{A} := \hat{A} - \bra \hat{A} \ket \hat{I} ,
	\\
&&	| \alpha \ket := \Delta \hat{A} | \psi \ket,
	\\
&&	| \beta \ket  := \Delta \hat{B} | \psi \ket,
\end{eqnarray}
where $ \hat{I} $ is the identity operator on $ {\mathscr H} $.
Then the Schwarz inequality (\ref{Schwarz}) becomes
\begin{equation}
	\bra \psi | (\Delta \hat{A})^2 | \psi \ket
	\bra \psi | (\Delta \hat{B})^2 | \psi \ket
	\ge
	\Big| \bra \psi | \Delta \hat{A} \, \Delta \hat{B} | \psi \ket \Big|^2.
	\label{rewrite Schwarz}
\end{equation}
The standard deviation of the observable $ \hat{A} $ is defined as
\begin{equation}
	\sigma( \hat{A} ) :=
	\Big(
	\bra \psi | (\Delta \hat{A})^2 | \psi \ket 
	\Big)^{\frac{1}{2}}.
\end{equation}
Then it is easy to see that
\begin{eqnarray}
	\Delta \hat{A} \, \Delta \hat{B} 
	&=&
	\frac{1}{2} 
	\Big( \Delta \hat{A} \, \Delta \hat{B} + \Delta \hat{B} \, \Delta \hat{A} \Big)
	+ \frac{1}{2}
	\Big( \Delta \hat{A} \, \Delta \hat{B} - \Delta \hat{B} \, \Delta \hat{A} \Big)
	\nonumber \\
	&=&
	  \frac{1}{2} \{ \Delta \hat{A}, \Delta \hat{B} \}
	+ \frac{1}{2} [ \Delta \hat{A}, \Delta \hat{B}  ].
\end{eqnarray}
Since
$ \bra \psi | \{ \Delta \hat{A}, \Delta \hat{B} \} | \psi \ket $
is a real number and
$ \bra \psi |  [ \Delta \hat{A}, \Delta \hat{B}  ] | \psi \ket $
is a pure imaginary number,
the right-hand side of (\ref{rewrite Schwarz}) can be rewritten as
\begin{eqnarray}
	\Big| \bra \psi | \Delta \hat{A} \, \Delta \hat{B} | \psi \ket \Big|^2
	=
	  \frac{1}{4} 
	\bra \psi | \{ \Delta \hat{A}, \Delta \hat{B} \} | \psi \ket^2
	+ \frac{1}{4} 
	\Big| \bra \psi |  [ \Delta \hat{A}, \Delta \hat{B}  ] | \psi \ket \Big|^2.
\end{eqnarray}
Moreover, we can see that
\begin{eqnarray}
	[ \Delta \hat{A}, \Delta \hat{B}  ] = [ \hat{A}, \hat{B} ].
\end{eqnarray}
Therefore, (\ref{rewrite Schwarz}) implies
\begin{eqnarray}
	\sigma ( \hat{A} )^2 
	\sigma ( \hat{B} )^2
& \ge &
	\Big| \bra \psi | \Delta \hat{A} \, \Delta \hat{B} | \psi \ket \Big|^2
	=
	  \frac{1}{4} 
	\bra \psi | \{ \Delta \hat{A}, \Delta \hat{B} \} | \psi \ket^2
	+ \frac{1}{4} 
	\Big| \bra \psi |  [ \Delta \hat{A}, \Delta \hat{B}  ] | \psi \ket \Big|^2
	\nonumber \\
& \ge &
	\frac{1}{4} \Big| \bra \psi |  [ \hat{A}, \hat{B} ] | \psi \ket \Big|^2.
	\label{rewrite Schwarz again}
\end{eqnarray}
By taking squre roots of the both side, we obtain the Robertson inequality
\begin{equation}
	\sigma (\hat{A} ) \cdot
	\sigma ( \hat{B} )
	\ge
	\frac{1}{2} \Big| \bra \psi |  [ \hat{A}, \hat{B} ] | \psi \ket \Big|,
	\label{Robertson}
\end{equation}
which means that the two observables cannot have precise values simultaneously
if $ \bra \psi |  [ \hat{A}, \hat{B} ] | \psi \ket $ $ \ne 0 $.
On the other hand, the quantity
\begin{eqnarray}
	C_s ( \hat{A}, \hat{B} )
& := &
	\frac{1}{2} 
	\bra \psi | \{ \Delta \hat{A}, \Delta \hat{B} \} | \psi \ket
	=
	\frac{1}{2} 
	\bra \psi | \{ \Delta \hat{A}, \hat{B} \} | \psi \ket
	=
	\frac{1}{2} 
	\bra \psi | \{ \hat{A}, \Delta \hat{B} \} | \psi \ket
	\nonumber \\
& = &
	\frac{1}{2} 
	\bra \psi | \{ \hat{A}, \hat{B} \} | \psi \ket
	-
	\bra \psi | \hat{A} | \psi \ket
	\bra \psi | \hat{B} | \psi \ket
	\label{covariance}
\end{eqnarray}
is called the symmetrized covariance of $ \hat{A} $ and $ \hat{B} $.
Then (\ref{rewrite Schwarz again}) can be rewritten as
\begin{eqnarray}
	\sigma ( \hat{A} )^2 \cdot
	\sigma ( \hat{B} )^2
& \ge &
	\Big| C_s ( \hat{A}, \hat{B} ) \Big|^2
	+ \frac{1}{4} 
	\Big| \bra \psi |  [ \hat{A}, \hat{B}  ] | \psi \ket \Big|^2.
	\label{Schrodinger inequality}
\end{eqnarray}
Sometimes this is referred 
as the Schr{\"o}dinger inequality \cite{Schrodinger1930}.

\section{Angular order parameter and orbital angular momentum}
In this section we show our main result.
Let us consider a quantum mechanical particle 
in a configuration space whose dimension is equal to or larger than two.
The system has four observables 
$ \hat{x}, \hat{y}, \hat{p}_x, \hat{p}_y $, which satisfy
the canonical commutation relations
$ [ \hat{x}_j, \hat{p}_k ] = i \hbar \delta_{jk} $.
We introduce two operators
\begin{eqnarray}
	\hat{Z} := \hat{x} + i\hat{y},
	\qquad
	\hat{L} := \hat{x} \hat{p}_y - \hat{y} \hat{p}_x.
\end{eqnarray}
The operator $ \hat{Z} $ is not self-adjoint
but it is related to position of the particle.
The self-adjoint operator $ \hat{L} $ 
is called the orbital angular momentum (OAM).
They satisfy
\begin{eqnarray}
	[ \hat{L}, \hat{Z} ] = \hbar \hat{Z}
	\label{LZ commutator}
\end{eqnarray}
and also
\begin{eqnarray}
	[ \hat{L}, \hat{Z}^n ] = n \hbar \, \hat{Z}^n
\end{eqnarray}
for any natural number $ n = 1, 2, 3, \cdots $.
With a normalized vector $ \psi \in {\mathscr H} $ we define
\begin{eqnarray}
	\bra \hat{L} \ket := \bra \psi | \hat{L} | \psi \ket,
	\qquad
	\Delta \hat{L} := \hat{L} - \bra \hat{L} \ket \hat{I}.
\end{eqnarray}
By substituting
\begin{eqnarray}
	| \alpha \ket = \Delta \hat{L} | \psi \ket,
	\qquad
	| \beta \ket = \hat{Z} | \psi \ket
\end{eqnarray}
into the Schwarz inequality (\ref{Schwarz}) and by noting
$ \bra \alpha | 
= \bra \psi | \Delta \hat{L}^\dagger
= \bra \psi | \Delta \hat{L} $
and
$ \bra \beta | = \bra \psi | \hat{Z}^\dagger $,
we get
\begin{equation}
	\bra \psi | (\Delta \hat{L})^2 | \psi \ket
	\bra \psi | \hat{Z}^\dagger \hat{Z} | \psi \ket
	\ge
	\Big| \bra \psi | \Delta \hat{L} \, \hat{Z} | \psi \ket \Big|^2.
	\label{rewrite Schwarz1}
\end{equation}
Hence 
\begin{equation}
	\sqrt{ \bra (\Delta \hat{L})^2 \ket }
	\sqrt{ \bra \hat{Z}^\dagger \hat{Z} \ket }
	\ge
	\Big| \bra \Delta \hat{L} \, \hat{Z} \ket \Big|.
	\label{rewrite Schwarz2}
\end{equation}
In a similar way, by substituting
\begin{eqnarray}
	| \alpha \ket = \hat{Z}^\dagger | \psi \ket,
	\qquad
	| \beta \ket = \Delta \hat{L} | \psi \ket
\end{eqnarray}
into (\ref{Schwarz}), we get
\begin{equation}
	\sqrt{ \bra \hat{Z} \hat{Z}^\dagger \ket }
	\sqrt{ \bra (\Delta \hat{L})^2 \ket }
	\ge
	\Big| \bra \hat{Z} \Delta \hat{L} \ket \Big|.
	\label{rewrite Schwarz3}
\end{equation}
Note that $ \hat{Z} \hat{Z}^\dagger = \hat{Z}^\dagger \hat{Z} $.
The triangle inequality 
$ |a| + |b| \ge | a-b | $ holds for any complex number $ a, b $.
The commutation relation (\ref{LZ commutator}) implies
$ [ \Delta \hat{L}, \hat{Z} ]
= [ \hat{L}, \hat{Z} ] = \hbar \hat{Z} $.
By adding (\ref{rewrite Schwarz2}) with (\ref{rewrite Schwarz3})
and multiplying $ 1/2 $,
we obtain
\begin{eqnarray}
	\sqrt{ \bra (\Delta \hat{L})^2 \ket }
	\sqrt{ \bra \hat{Z}^\dagger \hat{Z} \ket }
	& \ge &
	\frac{1}{2}
	\Big\{
		\big| \bra \Delta \hat{L} \, \hat{Z} \ket \big| +
		\big| \bra \hat{Z} \Delta \hat{L} \ket \big|
	\Big\}
	\nonumber \\
	& \ge &
	\frac{1}{2}
	\Big\{
	\big| \bra \Delta \hat{L} \hat{Z} - \hat{Z} \Delta \hat{L} \ket \big|
	\Big\}
	\nonumber \\
	& = &
	\frac{1}{2} \hbar \,
	\big| \bra \hat{Z} \ket \big|.
	\label{result1}
\end{eqnarray}
This is one of our main results.

By replacing the operator $ \hat{Z} $ with $ \hat{Z}^n $
we can derive more general inequalities
\begin{eqnarray}
	\sqrt{ \bra (\Delta \hat{L})^2 \ket }
	\sqrt{ \bra ( \hat{Z}^\dagger \hat{Z} )^n \ket }
	& \ge &
	\frac{1}{2} n \hbar \,
	\big| \bra \hat{Z}^n \ket \big|
	\qquad
	( n = 1, 2, 3, \cdots )
	\label{general UR}
\end{eqnarray}
via a similar inference. The nonnegative number
\begin{eqnarray}
	\sigma( \hat{L} )
	:= \sqrt{ \bra \psi | (\Delta \hat{L})^2 | \psi \ket }
	= \bra \psi | \hat{L}^2 | \psi \ket - \bra \psi | \hat{L} | \psi \ket^2
\end{eqnarray}
is the standard deviation of the orbital angular momentum.
The complex number 
\begin{eqnarray}
	\bra \hat{Z}^n \ket 
	= \bra \psi | (  \hat{x}+i \hat{y})^n | \psi \ket
	= \int \!\!\! \int_{- \infty}^{\infty}
	( x+iy )^n \, \Big| \psi(x,y) \Big|^2 \, dx \, dy
	\label{n-th moment}
\end{eqnarray}
is the {\it $ n $-th moment} of probability density
for the wave function $ \psi(x,y) $\footnote{
If the dimension of the configuration space is larger than two,
it is necessary to use a suitable wave function $ \psi ( x,y,x, \cdots ) $. }.
If the probability density $ | \psi(x,y) |^2 $ is rotationally invariant,
all the moment vanish $ \bra \hat{Z}^n \ket = 0 $ $ (n=1,2,3, \cdots ) $.
By contraposition,
if the system exhibits a nonvanishing moment $ \bra \hat{Z}^n \ket \ne 0 $
for some $ n $,
the probability density $ | \psi(x,y) |^2 $ is not rotationally invariant.
Hence, the expectation value $ \bra \hat{Z}^n \ket $ is interpreted
as an order parameter 
to measure the degree of breaking of the rotational symmetry.
The complex number
\begin{eqnarray}
	\mu_n :=
	\frac{ \bra \hat{Z}^n \ket }
	{ \sqrt{ \bra ( \hat{Z}^\dagger \hat{Z} )^n \ket } \; }
	=
	\frac{ \bra ( \hat{x}+i \hat{y})^n \ket }
	{ \sqrt{ \bra ( \hat{x}^2 + \hat{y}^2 )^n \ket } \; }
	\label{normalized n-th moment}
\end{eqnarray}
is called the {\it normalized $ n $-th moment of position distribution}
or the {\it normalized angular order parameter},
which indicates bias or asymmetry of angular distribution of the particle.
Then we have 
\begin{eqnarray}
	\sigma ( \hat{L} )
	\: \ge \:
	\frac{1}{2} n \hbar \, 
	\frac{ \: \big| \bra \hat{Z}^n \ket \big| \: }%
	{ \: \bra ( \hat{Z}^\dagger \hat{Z} )^n \ket^{1/2} \: }
	=
	\frac{1}{2} n \hbar \, \Big| \mu_n \Big|
	\qquad
	( n = 1, 2, 3, \cdots ).
	\label{result2}
\end{eqnarray}
This is the main result of our work.
This inequality implies that
if the uncertainty $ \sigma ( \hat{L} ) $ of OAM is small, 
the normalized moment $ | \mu_n | $ must be small.
In this case, the angular distribution is not strongly biased 
and hence the uncertainty of angle must be large.

Oppositely,
if the uncertainty of angle is small,
the angular distribution is strongly biased
and hence the normalized moment $ | \mu_n | $ becomes large,
then the inequality (\ref{result2}) implies that
the uncertainty $ \sigma ( \hat{L} ) $ of OAM must become large.

\section{Tighter inequality}
The necessary and sufficient conditions
for the equality in (\ref{result1}) are
the two equalities in 
(\ref{rewrite Schwarz2}), (\ref{rewrite Schwarz3})
and the other equality
$ \bra \Delta \hat{L} \hat{Z} \ket 
= - \bra \hat{Z} \Delta \hat{L} \ket $.
Actually, there is no state vector satisfying 
these three conditions simultaneously,
and hence the equality in (\ref{result1}) is never attained.
In this sense, the inequality (\ref{result1}) is not tight.

It is desirable to find a tighter inequality.
For this purpose, we introduce self-adjoint operators
\begin{equation}
	\hat{x}_n :=
	\frac{1}{2} ( \hat{Z}^n + \hat{Z}^{\dagger n} )
	\qquad
	\hat{y}_n :=
	\frac{1}{2i} ( \hat{Z}^n - \hat{Z}^{\dagger n} )
\end{equation}
for $ n = 1, 2, 3, \cdots $. Then we have
\begin{equation}
	\hat{Z}^n
	= \big( \hat{x} + i \hat{y} \big)^n
	= \hat{x}_n +i \hat{y}_n .
\end{equation}
Using these, it is easy to see that
\begin{eqnarray}
	\Delta \hat{L} \, \hat{Z}^n 
&=&
	\frac{1}{2} \{ \Delta \hat{L}, \hat{Z}^n \}
	+ \frac{1}{2} [ \Delta \hat{L}, \hat{Z}^n ]
	\nonumber \\
&=&
	\frac{1}{2} \{ \Delta \hat{L}, 
	( \hat{x}_n +i \hat{y}_n ) \} 
	+ \frac{1}{2} n \hbar \hat{Z}^n
	\nonumber \\
&=&
	\frac{1}{2} \{ \Delta \hat{L}, \hat{x}_n \} 
	+i \frac{1}{2} \{ \Delta \hat{L}, \hat{y}_n \} 
	+ \frac{1}{2} n \hbar ( \hat{x}_n +i \hat{y}_n ).
	\label{rewrite}
\end{eqnarray}
Hence, (\ref{rewrite Schwarz1}) is equivalent to
\begin{eqnarray}
	\bra (\Delta \hat{L})^2 \ket \cdot
	\bra \hat{Z}^\dagger \hat{Z} \ket 
& \ge &
	\Big| 
	\frac{1}{2} \bra \{ \Delta \hat{L}, \hat{x}_n \} \ket
	+ \frac{1}{2} n \hbar \bra \hat{x}_n \ket 
	\Big|^2
	+ 
	\Big| 
	\frac{1}{2} \bra \{ \Delta \hat{L}, \hat{y}_n \} \ket
	+ \frac{1}{2} n \hbar \bra \hat{y}_n \ket 
	\Big|^2
	\nonumber \\
& = &
	\Big| 
	C_s ( \hat{L}, \hat{x}_n )
	+ \frac{1}{2} n \hbar \bra \hat{x}_n \ket \Big|^2
	+ 
	\Big| 
	C_s ( \hat{L}, \hat{y}_n )
	+ \frac{1}{2} n \hbar \bra \hat{y}_n \ket \Big|^2.
	\label{result3}
\end{eqnarray}
This is the tightest inequality whose equality can be attained.
However, the equality holds
if and only if the state is an eigenstate of $ \hat{L} $.
In this case the both sides of (\ref{result3}) are zero.

\section{Experimental realization}
We have formulated the uncertainty relations
(\ref{result1}),
(\ref{result2}),
(\ref{result3}).
For testing these relations
we need to have a method 
for controlling and measuring angular momenta of particles.

In optics there is a method 
for controlling and measuring angular momenta of photons.
Franke-Arnold and Padgett {\it et al.}~\cite{Franke-Arnold 2004, Franke-Arnold 2005}
have tested the uncertainty relation of
Judge~\cite{Judge1963} and Berbett, Pegg~\cite{Bernett-Pegg1990},
by using an analyzer of photon angular momentum.

Uchida and Tonomura \cite{Uchida2010}
first made a coherent electron beam
carrying nonzero orbital angular momentum.
Such electron beam has a wave front whose shape looks like a vortex.
Verbeeck {\it et al.} \cite{Verbeeck2010} and
McMorran {\it et al.} \cite{McMorran2011}
developed fork-shaped diffraction gratings,
which control orbital angular momenta of electrons.
They observed circularly symmetric diffraction patterns
for eigenstates of orbital angular momentum.
Thus they verified that
the uncertainty in angular distribution was maximum
when the uncertainty of angular momentum was minimum.

Recently, Hasegawa and Saitoh {\it et al.} 
\cite{Saitoh2013a,Saitoh2013b}
made superposition of
two coherent electron beams carrying different angular momenta.
As a result, they produced a quantum state 
that has an uncertain orbital angular momentum.
They observed an interference pattern that was circularly asymmetric.
Thus they verified that
the uncertainty in angular distribution became smaller
when the uncertainty of angular momentum became larger.
Yet quantitative analysis of the uncertainty relation 
is not performed in experiments using electrons.

\section{Generalization}
The angular momentum $ \hat{L} $
is a generator of rotational transformations,
which transform the angle variable 
$ ( \hat{x} + i \hat{y} ) / \sqrt{ \hat{x}^2 + \hat{y}^2 } $.
A nonzero value of the order parameter 
$ \mu = \bra \hat{x} + i \hat{y} \ket 
/ \sqrt{ \bra \hat{x}^2 + \hat{y}^2 \ket } $
indicates breaking of rotational symmetry,
or certainty of the angle distribution,
which accompanies uncertainty of the angular momentum.
The relation between the angle order parameter 
and the uncertainty of the angular momentum
is expressed by the inequality (\ref{result2}).

This kind of relation between a symmetry generator
and a symmetry-breaking order parameter can be formulated
in a more general form.
Suppose that we have a self-adjoint operator $ \hat{G} $,
which is a generator of symmetry transformations
and is called a charge.
And suppose that we have some operator $ \hat{\varPhi} $.
It is not necessary to assume that 
$ \hat{\varPhi} $ is a self-adjoint operator.
Then the inequality
\begin{equation}
	\sigma ( \hat{G} )
	\: \ge \:
	\frac{ \: | \, \bra [ \hat{G}, \hat{\varPhi} ] \ket \, | \:}%
	{ \: \sqrt{ \bra \hat{\varPhi}^\dagger \hat{\varPhi} \ket }
	+ \sqrt{ \bra \hat{\varPhi} \hat{\varPhi}^\dagger \ket } \: }
	\label{result4}
\end{equation}
holds. 
The expectation value 
$ \bra [ \hat{G}, \hat{\varPhi} ] \ket
= \bra \psi | [ \hat{G}, \hat{\varPhi} ] | \psi \ket $
is taken with respect to a state $ | \psi \ket $.
This is a generalization of (\ref{result1})
and its proof is straightforward.

In the left-hand side of (\ref{result4})
the standard deviation $ \sigma ( \hat{G} ) $ measures
uncertainty of the charge.
In the right-hand side of (\ref{result4})
the commutator $  [ \hat{G}, \hat{\varPhi} ]  $ represents
transformation of $ \hat{\varPhi} $ by $ \hat{G} $.
If the state $ | \psi \ket $ is invariant 
under the transformation generated by $ \hat{G} $,
then 
$ \bra \psi | [ \hat{G}, \hat{\varPhi} ] | \psi \ket = 0 $.
If the order parameter 
$ \bra [ \hat{G}, \hat{\varPhi} ] \ket  $
exhibits a nonzero value,
then the state is not invariant
and the uncertainty of the charge must satisfy
the inequality (\ref{result4}).

This formulation is applicable to 
the uncertainty relation
between the particle number and the phase.
In this case we take
$ \hat{G} = \hat{a}^\dagger \hat{a} $
and $ \hat{\varPhi} = \hat{a} $,
with the creation and annihilation operators
$ \hat{a}^\dagger $ and $ \hat{a} $.

This formulation is applicable also to 
the complementarity relation \cite{tanimura2007}
between the particle nature and the wave nature.

\section{Summary}
The uncertainty relation
between angle and orbital angular momentum
does not have a formulation similar 
to the uncertainty relation
between position and linear momentum.
The angle variable is not represented
by a quantum mechanical self-adjoint operator
although the other observables are represented
by self-adjoint operators.
We reviewed the general formulation 
of the uncertainty relation between noncommutative observables,
which was proved by Robertson.
Instead of the angle variable operator,
we introduced the complex position operator 
$ \hat{Z} = \hat{x}+i \hat{y} $
and interpreted the order parameter
$ \mu = 
\bra \hat{Z} \ket / \sqrt{ \bra \hat{Z}^\dagger \hat{Z} \ket} $
as a measure of certainty of angle distribution.
Then we proved the relation (\ref{result1})
between the uncertainty of angular momentum
and the certainty of angle.
We proved the relations (\ref{result2}),
which are generalizations to higher moments of angular distribution
$ \mu_n = 
\bra \hat{Z}^n \ket 
/ \sqrt{ \bra (\hat{Z}^\dagger \hat{Z})^n \ket} $.
We proved also the tightest inequality (\ref{result3}).
A theoretical generalization 
to the uncertainty relation (\ref{result4})
between a symmetry generator and an order parameter 
was shown.
Methods for controlling angular momenta of photons and electrons were discussed.
Quantitative experimental tests of the relations 
are postponed for future work.

In this paper we considered uncertainties of values of observables
that are inherent in quantum states.
However we did not consider measurement process of observables.
An actual measurement process involves measurement error and 
causes disturbance on the state of the measured system.
Ozawa \cite{Ozawa2003} formulated a quantitative relation
between the measurement error and the disturbance.
Branciard \cite{Branciard2013} established the tightest inequality
that the error and the disturbance obey.
We do not yet know this kind of error-disturbance relation
for the angle and the angular momentum.

Hayashi \cite{Hayashi2012} formulated 
quantum estimation theory for the group action,
which can be regarded as a generalization
of the problem that was considered in our work.
This aspect should be investigated more.

\section*{Acknowledgements}
The author thanks Keisuke Watanabe, 
who told me the tighter version of the uncertainty inequality, 
Eq. (\ref{result3}).
He thanks
Prof. Katsuhiro Nakamura and
Prof. Davron Matrasulov
for their warm hospitality for supporting his stays in Uzbekistan.
This manuscript is written as a part of the proceedings of the workshop,
Wave dynamics in low-dimensional branched structures,
held during September 23-24, 2014 in Tashkent, Uzbekistan.
This work is financially supported 
by the Grant-in-Aid for Scientific Research
of Japan Society for the Promotion of Science,
Grant No.~26400417.


\begin{thebibliography}{99}
\baselineskip 5mm

\bibitem{Heisenberg1927}
Heisenberg, W.
``{\"U}ber den anschaulichen Inhalt der quantentheoretischen Kinematik und Mechanik,''
Z. Physik {\bf 43}, 172 (1927).

\bibitem{Robertson1929}
Robertson, H. P.
``The uncertainty principle,''
Phys. Rev. {\bf 34}, 163 (1929).

\bibitem{Schiff1968}
Schiff, L. I. 
{\it Quantum Mechanics}, 3rd edition (McGraw-Hill, 1968).
In Eq. (3.2) of the textbook the naive uncertainty relation
$ \Delta \phi \cdot \Delta J_z \gtrsim \hbar $ is shown.

\bibitem{Judge1963}
Judge, D. 
``On the uncertainty relation for $ L_z $ and $ \varphi $,''
Phys. Lett. {\bf 5}, 189 (1963).

\bibitem{Kraus1965}
Kraus, K. 
``Remark on the uncertainty between angle and angular momentum,'' 
Z. Physik {\bf 188}, 374 (1965).

\bibitem{Carruthers1968}
Carruthers, P.,
Nieto, M. M.
``Phase and angle variables in quantum mechanics,''
Rev. Mod. Phys. {\bf 40}, 411 (1968).

\bibitem{Bernett-Pegg1990}
Bernett, S. M., 
Pegg, D. T. 
``Quantum theory of rotation angles,''
Phys. Rev. A {\bf 41}, 3427 (1990).

\bibitem{Ohnuki-Kitakado1993}
Ohnuki, Y., Kitakado, S. 
``Fundamental algebra for quantum mechanics on $ S^D $ and gauge potentials,''
J. Math. Phys. {\bf 34}, 2827 (1993).

\bibitem{Tanimura1993}
Tanimura, S. 
``Gauge field, parity and uncertainty relation of quantum mechanics on $ S^1 $,''
Prog. Theor. Phys. {\bf 90}, 271 (1993).

\bibitem{Schrodinger1930}
Schr{\"o}dinger, E. 
``Zum Heisenbergschen Unsch{\"a}rfeprinzip,''
Sitzungsberichte der Preussischen Akademie der Wissenschaften,
Physikalisch-mathematische Klasse {\bf 19}, 296 (1930).

\bibitem{Franke-Arnold 2004}
Franke-Arnold, S.,
Barnett, S. M.,
Yao, E.,
Leach, J.,
Courtial, J.,
Padgett, M. 
``Uncertainty principle for angular position and angular momentum,''
New J. Phys. {\bf 6}, 103 (2004).

\bibitem{Franke-Arnold 2005}
Pegg, D. T.,
Barnett, S. M.,
Zambrini, R.,
Franke-Arnold, S.,
Padgett, M.
``Minimum uncertainty states of angular momentum and angular position,''
New J. Phys. {\bf 7}, 62 (2005).

\bibitem{Uchida2010}
Uchida, M.
Tonomura, A.
``Generation of electron beams carrying orbital angular momentum,''
Nature {\bf 464}, 737 (2010).

\bibitem{Verbeeck2010}
Verbeeck, J.,
Tian, H.,
Schattschneider, P.
``Production and application of electron vortex beams,''
Nature {\bf 467}, 301 (2010).

\bibitem{McMorran2011}
McMorran, B. J.,
Agrawal, A.,
Anderson, I. M.,
Herzing, A. A.,
Lezec, H. J.,
McClelland, J. J.,
Unguris, J. 
``Electron vortex beams with high quanta of orbital angular momentum,''
Science {\bf 331}, 192 (2011).

\bibitem{Saitoh2013a}
Hasegawa, Y.,
Saitoh, K.,
Tanaka, N.,
Tanimura, S.,
Uchida, M.
``Young's interference experiment with electron beams carrying orbital angular momentum,''
J. Phys. Soc. Jap. {\bf 82}, 033002 (2013).

\bibitem{Saitoh2013b}
Hasegawa, Y.,
Saitoh, K.,
Tanaka, N.,
Uchida, M.
``Propagation dynamics of electron vortex pairs,''
J. Phys. Soc. Jap. {\bf 82}, 073402 (2013).

\bibitem{tanimura2007}
Tanimura, S.,
``The incompatibility relation between
visibility of interference and distinguishability of paths,''
e-print arXiv quant-ph/0703118 (2007).

\bibitem{Ozawa2003}
Ozawa, M.,
``Universally valid reformulation of the Heisenberg uncertainty principle on noise and disturbance in measurement,''
Phys. Rev. A {\bf 67}, 042105 (2003).

\bibitem{Branciard2013}
Branciard, C.,
``Error-tradeoff and error-disturbance relations for incompatible quantum measurements,''
Proceedings of the National Academy of Science 
of the United States of America {\bf 110}, 6742 (2013).
Supporting Information is also available.

\bibitem{Hayashi2012}
Hayashi, M.,
``Fourier analytic approach to quantum estimation of group action,''
e-print arXiv 1209.3463v2 (2012).

\end{thebibliography}
\end{document}